\documentclass[manuscript]{aastex}

\shorttitle{Debris disks and planets}
\shortauthors{K\'osp\'al et al.}

\begin{document}


\title{On the Relationship Between Debris Disks and Planets}

\author{\'Agnes K\'osp\'al}
\affil{Leiden Observatory, Leiden University,\\ P.O. Box 9513,
2300 RA, Leiden, The Netherlands}
\email{kospal@strw.leidenuniv.nl}

\author{David R. Ardila}
\affil{NASA Herschel Science Center, California Institute of Technology,\\
Pasadena, CA 91125, USA}

\and

\author{Attila Mo\'or and P\'eter \'Abrah\'am}
\affil{Konkoly Observatory of the Hungarian Academy of Sciences,\\ P.O.
Box 67, 1525 Budapest, Hungary}


\begin{abstract}
  Dust in debris disks is generated by collisions among
  planetesimals. The existence of these planetesimals is a consequence
  of the planet formation process, but the relationship between debris
  disks and planets has not been clearly established. Here we analyze
  Spitzer/MIPS 24 and 70$\,\mu$m data for 150 planet-bearing stars,
  and compare the incidence of debris disks around these stars with a
  sample of 118 stars around which planets have been searched for, but
  not found. Together they comprise the largest sample ever assembled
  to deal with this question. The use of survival analysis techniques
  allows us to account for the large number of non-detections at
  70$\,\mu$m. We discovered 10 new debris disks around stars with
  planets and one around a star without known planets. We found that
  the incidence of debris disks is marginally higher among stars with
  planets, than among those without, and that the brightness of the
  average debris disk is not significantly different in the two
  samples. We conclude that the presence of a planet that has been
  detected via current radial velocity techniques is not a good
  predictor of the presence of a debris disk detected at infrared
  wavelengths.
\end{abstract}



\section{Introduction}
\label{sec:intro}

As of the end of 2008, 228 extrasolar planets around 193 stars had
been discovered using radial velocity techniques. Most of them are
believed to be Jupiter-like gas giants, although some may be
lower-mass rocky planets \citep{udry}. According to the core accretion
model, planets are formed by the agglomeration of rocky planetesimals
into solid cores \citep{lissauer}. For gas giants, the core accretes a
gaseous envelope. The planets may migrate toward the star due to
interactions with the disk material \citep{lin}, and strong
planet--planet gravitational scattering may significantly modify
orbital parameters \citep{raymond}.

It is also known that many main-sequence stars display infrared and/or
submillimeter excess emission due to the presence of debris dust
grains \citep[e.g.,][]{trilling2008}. These grains are replenished by
collisions among planetesimals, leftovers of the planet formation
process (Backman \& Paresce 1993). Planets orbiting the star can stir
the planetesimal population and initiate a collisional cascade,
resulting in dust production. The morphology of the disk might be
gravitationally shaped by planets orbiting interior to the disk
\citep{chiang}. The common origin of planets and debris disks suggests
that a connection may exist between the presence of planets and
planetesimals.

The relationship between debris disks and planets can be studied by
comparing the incidence of excess emission due to debris dust between
samples of stars with and without planets. An overview of such studies
is presented in Tab.~\ref{tab:detection}. For Sun-like stars
e.g.~\citet{beichman2005,beichman2006} hinted at a positive
correlation between debris disks and planets; \citet{moromartin} found
no significant correlation; and \citet{trilling2008} claimed that
although excess rates for the planet sample are higher, both samples'
excess rates are consistent at the 1\,$\sigma$ level. More conclusive
results can only be based on larger samples than used in those
studies.

Here we use data from the {\it Spitzer Space Telescope} archive at 24
and 70$\,\mu$m to compare the incidence of debris dust around stars
with planets (SWPs) discovered/confirmed using radial velocity
techniques and those where radial velocity measurements did not
indicate planetary companions (stars without planets, SWOPs). We use
all public data constructing the largest sample ever considered in
attacking this question.

\begin{deluxetable}{lccccc}
\tabletypesize{\scriptsize}
\tablecaption{{\it Spitzer}-Based Surveys of Debris Disks Around
  Main-Sequence Stars (Stars with Excess / Total Observed
  Sample).\label{tab:detection}}
\tablewidth{0pt}
\tablehead{
\colhead{Reference} & \colhead{Sample} & \multicolumn{2}{c}{24$\,\mu$m} &
\multicolumn{2}{c}{70$\,\mu$m}\\
                                &           & \colhead{SWP} & \colhead{SWOP} & \colhead{SWP}  & \colhead{SWOP}
}
\startdata
\citet{beichman2005}            & FGK       & 0/25 (0\%)         & \nodata               & 6/25 (24\%)         & \nodata               \\
\citet{bryden2006}              & FGK       & 1/12 (8\%)         & 0/57 (0\%)            & 1/12 (8\%)          & 6/57 (10\%)           \\
\citet{beichman2006}            & FGKM      & \nodata            & 5/88 (6\%)            & \nodata             & 12/88 (14\%)          \\
\citet{su2006}                  & A         & \nodata            & 50/155 (32\%)         & \nodata             & 44/134 (33\%)         \\
\citet{gautier2007}             & M         & 0/3 (0\%)          & 0/59 (0\%)            & 0/3 (0\%)           & 0/38 (0\%)            \\
\citet{moromartin}              & FGK       & 0/9 (0\%)          & 2/99 (2\%)            & 1/9 (11\%)          & 9/99 (9\%)            \\
\citet{trilling2008}            & FGK       & 3/45 (7\%)         & 4/139 (3\%)           & 10/48 (21\%)        & 22/148 (15\%)         \\
\citet{carpenter2009}           & FGK       & 0/10 (0\%)         & 41/299 (14\%)         & 2/10 (20\%)         & 19/299 (6\%)          \\
{\bf present work}              &{\bf FGKM} & {\bf 3/83 (4\%)}   & {\bf 1/118 (1\%)}     & {\bf 22/150 (15\%)} & {\bf 17/117 (15\%)}   \\
\enddata
\end{deluxetable}


\section{Observations}
\label{sec:data}

\subsection{Sample selection}

We considered 193 SWPs discovered/confirmed using radial velocity
measurements as of 2008 December\footnote{Regularly updated lists of
  extrasolar planets can be found at
  \texttt{http://www.exoplanets.org} and at
  \texttt{http://www.exoplanet.eu}.}. Out of these, 150 have publicly
available {\it Spitzer}/MIPS 24 and/or 70$\,\mu$m measurements (all of
them were observed at 70$\,\mu$m, 83 of them at 24$\,\mu$m). These
constitute our planet-bearing sample. For SWOPs, we used those stars
that were present in the Keck, Lick, and Anglo-Australian Telescope
planet search programs \citep{wright, vf}, but around which no planets
were found. These stars have no planets with radial velocity
semiamplitudes $K > 30\,$ms$^{-1}$ and orbital periods shorter than
four years, and the radial velocity measurements have a typical
Doppler precision of 3 ms$^{-1}$. We based our SWOPs sample on two
Spitzer surveys of nearby stars (Pid 41 and Pid 2324) and discarded
some stars where close-by objects would have contaminated the MIPS
photometry. The resulting SWOPs sample includes 118 objects (all of
them were observed at 24$\,\mu$m, all but one at 70$\,\mu$m, see
Tab.~\ref{tab:data}). Both samples contain stars with spectral types
between F3 and M3. While the majority are main-sequence stars, a few
subgiants are also present, and the SWPs sample also contains four
giants. As age is a key parameter in debris disk evolution
\citep{wyatt2008}, we checked whether the ages differed significantly
in the two samples. We calculated two-sample tests (Gehan's
generalized Wilcoxon test, logrank test, Peto \& Peto generalized
Wilcoxon test, and Peto \& Prentice generalized Wilcoxon test), and
found that the age distributions do not differ (mean ages are
5.317\,Gyr for SWPs and 4.648\,Gyr for SWOPs). The same holds for the
effective temperatures (mean effective temperatures are 5460\,K for
SWPs and 5550\,K for SWOPs).


\subsection{Data Processing}

\paragraph{Data Reduction} From the Spitzer archive we downloaded BCD
files reduced with the SSC pipeline version 16.0 or 16.1. At 70
$\mu$m, we used GeRT (version S14.0 v1.1 [060415]) to do column mean
subtraction and time filtering on the BCD files. We used Mopex
(version 18.1.5) to create mosaics for 24 and 70$\,\mu$m images and
obtained aperture photometry using IDL. At 24$\,\mu$m, we used an
aperture radius of 7$''$, sky annulus between 40$''$ and 50$''$, and
aperture correction of 1.61; at 70$\,\mu$m, we used an aperture radius
of 8$''$, sky annulus between 39$''$ and 65$''$, and aperture
correction of 3.70, appropriate for a 10,000\,K blackbody.  After
identifying stars with 70$\,\mu$m excess, we recalculated their fluxes
with an aperture correction of 3.83, appropriate for a 60\,K
blackbody. Appropriate color corrections were also performed. Thanks
to the small aperture, nearby sources had no effect on our
photometry. At 24$\,\mu$m, the aperture was placed around the centroid
of each star, while at 70$\,\mu$m, the aperture was placed at the same
position as for the 24$\,\mu$m image. If a 24$\,\mu$m position was not
available, we used fixed coordinates. For sources brighter than
100\,mJy and having signal-to-noise ratio of $\geq$11, we checked
their spatial extent by comparing the images to a stellar point-spread
function (PSF). For extended objects (HD\,10647, HD\,38858,
HD\,207129, HD\,48682, and HD\,115617) photometry was extracted by
fitting a PSF broadened by an appropriate two-dimensional
Gaussian. The fluxes of $\epsilon$ Eri were taken from
\citet{backman}.

\paragraph{Photosphere Prediction} We collected near-infrared
photometry from the 2MASS All-Sky Catalog of Point Sources
\citep{2mass}, \citet{morel}, and \citet{ducati}, 8.28$\,\mu$m fluxes
from the MSX Infrared Point-Source Catalog \citep{msx}, and 12$\,\mu$m
fluxes from the IRAS catalog of point sources (with appropriate
color correction). Spectral types, effective temperatures, and
metallicities came from the NASA Star and Exoplanet
Database\footnote{\texttt{http://nsted.ipac.caltech.edu/}}. For each
FGK star we selected the model from the ATLAS9 grid of model
atmospheres \citep{kurucz} that has the closest metallicity, log
$g$=4.5, and interpolated between the two closest models in effective
temperature. This model spectrum was then scaled to the near- and
mid-infrared photometric points, and its value at 23.68$\,\mu$m and at
71.42$\,\mu$m was calculated. M stars were similarly fitted with
NextGen model atmospheres \citep{nextgen}. The resulting
  photospheric fluxes ($F^{\rm pred}_{24}$, $F^{\rm pred}_{70}$) are
presented in Tab.~\ref{tab:data}.

\paragraph{Empirical Correction} To test our photometric predictions,
we considered the ratio of the observed and predicted flux ($R_{24}$,
$R_{70}$) as a function of the logarithm of the observed brightness
($\log(F^{\rm obs}_{24})$, $\log(F^{\rm obs}_{70})$). We found that
$F^{\rm obs}_{24}$ tends to be systematically higher than $F^{\rm
  pred}_{24}$ for bright stars. At 70$\,\mu$m no such trend could be
seen. At 24$\,\mu$m the points could be well fitted with a linear
relationship, which we used to correct the observed fluxes (note that
\citealt{gordon} found a similar dependence of the 70$\,\mu$m aperture
photometry on source brightness, see their Figures 3 and 4). The
correction was always less than 6\%. We found no dependence of the
corrected flux on the effective temperature. This is in agreement with
\citet{engelbracht}, though is in contrast with
\citet{beichman2006}. Our {\it Spitzer}/MIPS 24 and 70$\,\mu$m
photometry, together with individual uncertainties (including those of
the absolute flux calibration of the MIPS instrument, 4\% at 24 and
7\% at 70$\,\mu$m, see the MIPS Data Handbook) can be seen in
Tab.~\ref{tab:data}.


\section{Results}
\label{sec:results}

\paragraph{Identification of Debris Disks} In all 24$\,\mu$m
observations the target was detected at $>3 \sigma$ level, where
$\sigma$ is the flux uncertainty. At 70$\,\mu$m, 59 out of 150 SWPs,
and 97 out of 117 SWOPs were detected at $>3 \sigma$ level. For the
non-detections at 70$\,\mu$m, upper limits were calculated as
$3\sigma_{70}$ if the observed flux in the aperture ($F^{\rm
  obs}_{70}$) was negative, and $F^{\rm obs}_{70} + 3\sigma_{70}$ if
positive \citep{carpenter2009}. A star has significant excess if
$\chi_{70}=(F^{\rm obs}_{70}-F^{\rm pred}_{70})/\sigma_{70}>3$
(Tab.~\ref{tab:data}). In the SWPs sample, three stars have significant
excess at 24$\,\mu$m, and 22 stars at 70$\,\mu$m. In the SWOPs sample,
one star shows significant 24$\,\mu$m excess, while 17 stars have
significant excess at 70$\,\mu$m. The names of stars with excess are
listed in the notes to Tab.~\ref{tab:data}.

\begin{figure}
\begin{center}
\includegraphics[angle=0,scale=1]{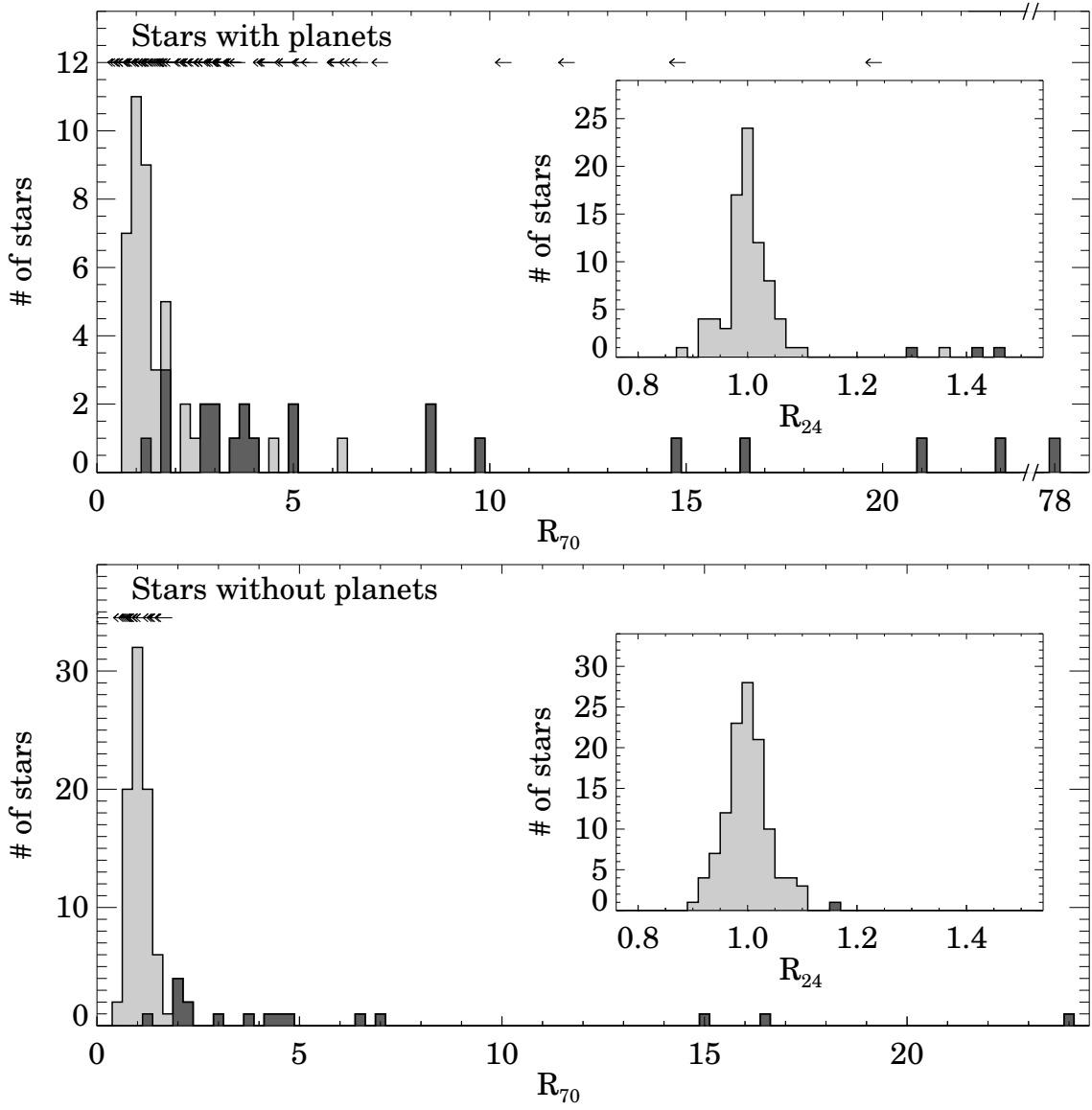}
\end{center}
\caption{Distribution of observed-to-predicted flux ratios at
  70 and 24$\,\mu$m ($R_{70}$ in the big graphs, and $R_{24}$
  in the small insets). Stars with significant (${>}\,$3$\,\sigma$)
  excess are indicated with dark shades. Upper limits are indicated
  with left arrows.
  \label{fig:hist}}
\end{figure}

Figure \ref{fig:hist} shows the histogram of observed-to-photospheric
flux ratios at 24 and 70$\,\mu$m (R$_{24}$, R$_{70}$). At both
wavelengths the distribution peaks at R$\,{\approx}\,$1, and can be
approximated with a Gaussian of $\sigma=4\%$ at 24$\,\mu$m (the error
budget is dominated by the uncertainty of the absolute flux
calibration) and $\sigma=24\%$ at 70$\,\mu$m (the photometry is
confusion noise limited at this wavelength). Out of the 39 stars with
excess, shown with dark shades in Figure \ref{fig:hist}, 28 were known
in the literature to harbour debris disks. For the remaining 11 our
work shows the first indication for a debris disk
(Tab.~\ref{tab:new}).

\begin{deluxetable}{@{}cc@{}c@{}c@{}c@{}c@{}c@{}c@{}c@{}c@{}c@{}c@{}c@{}c@{}c@{}cc}
\tabletypesize{\scriptsize}
\rotate
\tablecaption{Photospheric Predictions and MIPS Photometry.\label{tab:data}}
\tablewidth{0pt}
\tablehead{
\colhead{Name} & \colhead{Type}& \colhead{Age}   & \colhead{Age} & \colhead{[Fe/H]} & \colhead{$F^{\rm pred}_{24}$} & \colhead{$F^{\rm obs}_{24}$} & \colhead{$\sigma_{24}$} & \colhead{$R_{24}$} & \colhead{$\chi_{24}$} & \colhead{IRE$_{24}$\tablenotemark{a}} & \colhead{$F^{\rm pred}_{70}$} & \colhead{$F^{\rm obs}_{70}$} & \colhead{$\sigma_{70}$} & \colhead{$R_{70}$} & \colhead{$\chi_{70}$} & \colhead{IRE$_{70}$\tablenotemark{b}} \\
               &               & \colhead{(Gyr)} & \colhead{ref} &                  & \colhead{(mJy)}               & \colhead{(mJy)}              & \colhead{(mJy)}         &                    &                       &                      & \colhead{(mJy)}               & \colhead{(mJy)}              & \colhead{(mJy)}         &                    &                       &                       }
\startdata
HD 142         & SWP           & 3.120           & 1             &  0.0875          & 109.5                         & 119.6                        & 4.90                    & 1.09               & 2.06                  & NO                   & 11.92                         & 18.16                       & 2.73                     & 1.52               & 2.28                  & NO        \\
HD 166         & SWOP          & 0.200           & 2             & -0.0600          & 134.6                         & 157.4                        & 6.45                    & 1.17               & 3.53                  & YES                  & 14.68                         & 102.9                       & 8.42                     & 7.01               & 10.5                  & YES       \\
HD 1237        & SWP           & 0.625           & 3             &  0.1100          & 79.40                         & 82.61                        & 3.39                    & 1.04               & 0.95                  & NO                   & 8.672                         & 11.07                       & 1.68                     & 1.28               & 1.43                  & NO        \\
HD 1581        & SWOP          & 4.840           & 1             & -0.2025          & 567.6                         & 527.1                        & 21.6                    & 0.93               & -1.87                 & NO                   & 62.13                         & 74.40                       & 7.93                     & 1.20               & 1.55                  & NO        \\
HD 3651        & SWP           & 8.200           & 4             & -0.0540          & 187.8                         & 188.2                        & 7.72                    & 1.00               & 0.05                  & NO                   & 20.51                         & 15.49                       & 5.12                     & 0.76               & -0.98                 & NO        \\
HD 3823        & SWOP          & 7.200           & 1             & -0.3100          & 110.8                         & 111.8                        & 4.58                    & 1.01               & 0.20                  & NO                   & 12.14                         & $<$13.49                    & \nodata                  & $<$1.11            & \nodata               & UPL       \\
HD 3795        & SWOP          & 11.600          & 1             & -0.6425          & 130.0                         & 135.7                        & 5.56                    & 1.04               & 1.03                  & NO                   & 14.30                         & 13.92                       & 2.49                     & 0.97               & -0.15                 & NO        \\
HD 4208        & SWP           & 7.760           & 1             & -0.2600          & 24.81                         & \nodata                      & \nodata                 & \nodata            & \nodata               & \nodata              & 2.722                         & $<$5.35                     & \nodata                  & $<$1.96            & \nodata               & UPL       \\
HD 4308        & SWP           & 8.680           & 1             & -0.3933          & 73.67                         & 74.73                        & 3.06                    & 1.01               & 0.35                  & NO                   & 8.081                         & $<$13.00                    & \nodata                  & $<$1.61            & \nodata               & UPL       
\enddata

\tablenotetext{a}{Infrared excess at 24$\,\mu$m. SWPs with excess: HD
  10647, HD 22049, HD 69830. SWOPs with excess: HD 166.}
\tablenotetext{b}{Infrared excess at 70$\,\mu$m. SWPs with excess: GJ
  581, HD 10647, HD 19994, HD 22049, HD 33636, HD 38529, HD 40979, HD
  46375, HD 50499, HD 50554, HD 52265, HD 69830, HD 73526, HD 82943,
  HD 117176, HD 128311, HD 137759, HD 178911, HD 187085, HD 192263, HD
  202206, HD 216435. SWOPs with excess: HD 166, HD 7570, HD 17925, HD
  20794, HD 20807, HD 22484, HD 30495, HD 38858, HD 43162, HD 48682,
  HD 72905, HD 76151, HD 115617, HD 118972, HD 158633, HD 206860, HD
  207129.}  \tablecomments{(This table is available in its entirety in
  a machine-readable form in the online journal. A portion is shown
  here for guidance regarding its form and content.)}
\end{deluxetable}

\begin{deluxetable}{lcccccc}
\tabletypesize{\scriptsize}
\tablecaption{Newly Identified Debris Disks.\label{tab:new}}
\tablewidth{0pt}
\tablehead{
\colhead{Name} & \colhead{$F_{70}^{\rm pred}$} & \colhead{$F_{70}^{\rm obs}$} & \colhead{$\sigma_{70}$} & \colhead{$R_{70}$} & \colhead{$\chi_{70}$} & \colhead{$L_{\rm dust}/L_*$} \\
\colhead{ }    & \colhead{(mJy)}               & \colhead{(mJy)}              & \colhead{(mJy)}         &                    & \colhead{ }           & \colhead{ }          
}
\startdata
GJ 581\tablenotemark{a}    & 5.98   & 16.47  &  2.94 &  2.753 & 3.567 & 1.2$\times 10^{-5}$ \\
HD 40979\tablenotemark{a}  & 5.11   & 14.24  &  2.73 &  2.784 & 3.346 & 1.2$\times 10^{-5}$ \\
HD 43162\tablenotemark{b}  & 10.01  & 19.82  &  2.95 &  1.980 & 3.329 & 1.0$\times 10^{-5}$ \\
HD 46375\tablenotemark{a}  & 3.23   & 74.68  & 19.04 & 23.102 & 3.753 & 2.3$\times 10^{-4}$ \\
HD 50499\tablenotemark{a}  & 3.62   & 10.70  &  2.16 &  2.957 & 3.273 & 1.3$\times 10^{-5}$ \\
HD 73526\tablenotemark{a}  & 0.91   & 19.11  &  5.38 & 20.943 & 3.382 & 1.7$\times 10^{-4}$ \\
HD 137759\tablenotemark{a} & 455.52 & 615.53 & 44.08 &  1.351 & 3.630 & 4.3$\times 10^{-6}$ \\
HD 178911\tablenotemark{a} & 2.23   & 11.43  &  1.88 &  5.119 & 4.891 & 3.6$\times 10^{-5}$ \\
HD 187085\tablenotemark{a} & 3.45   & 13.34  &  2.81 &  3.867 & 3.520 & 1.8$\times 10^{-5}$ \\
HD 202206\tablenotemark{a} & 2.00   & 29.51  &  3.77 & 14.735 & 7.290 & 1.1$\times 10^{-4}$ \\
HD 216435\tablenotemark{a} & 10.88  & 41.74  &  4.04 &  3.836 & 7.636 & 2.0$\times 10^{-5}$ \\
\enddata
\tablenotetext{a}{Stars with planets}
\tablenotetext{b}{Stars without planets}
\end{deluxetable}

\paragraph{Survival Analysis} At 24$\,\mu$m, all observed stars were
detected at several $\sigma$ level. Thus, the distribution of excesses
can be compared with traditional two-sample tests, which show that
SWPs and SWOPs do not differ. This is due to the fact that the
overwhelming majority of stars exhibit pure photospheric emission at
24$\,\mu$m.

The 70$\,\mu$m data set is censored---there are upper limits---and
survival analysis is required to compare the two distributions
\citep[e.g.,][]{moromartin}. The Kaplan-Meier (KM) estimator gives the
cumulative distribution of a statistical variable taking into account
the upper limits \citep[e.g.,][]{fn}. To use survival analysis, the
censoring should not depend on the variable itself. This condition can
be fulfilled by using the observed-to-photospheric flux ratio
($R_{24}$ and $R_{70}$) as the variable. In average SWPs are farther
than SWOPs. As a consequence, the former tend to be apparently fainter
than the latter, resulting in more non-detections in the SWPs sample,
introducing Malmquist bias. This bias is eliminated by using the
distance-independent observed-to-photospheric flux ratio.

\begin{figure}
\begin{center}
\includegraphics[angle=90,scale=1]{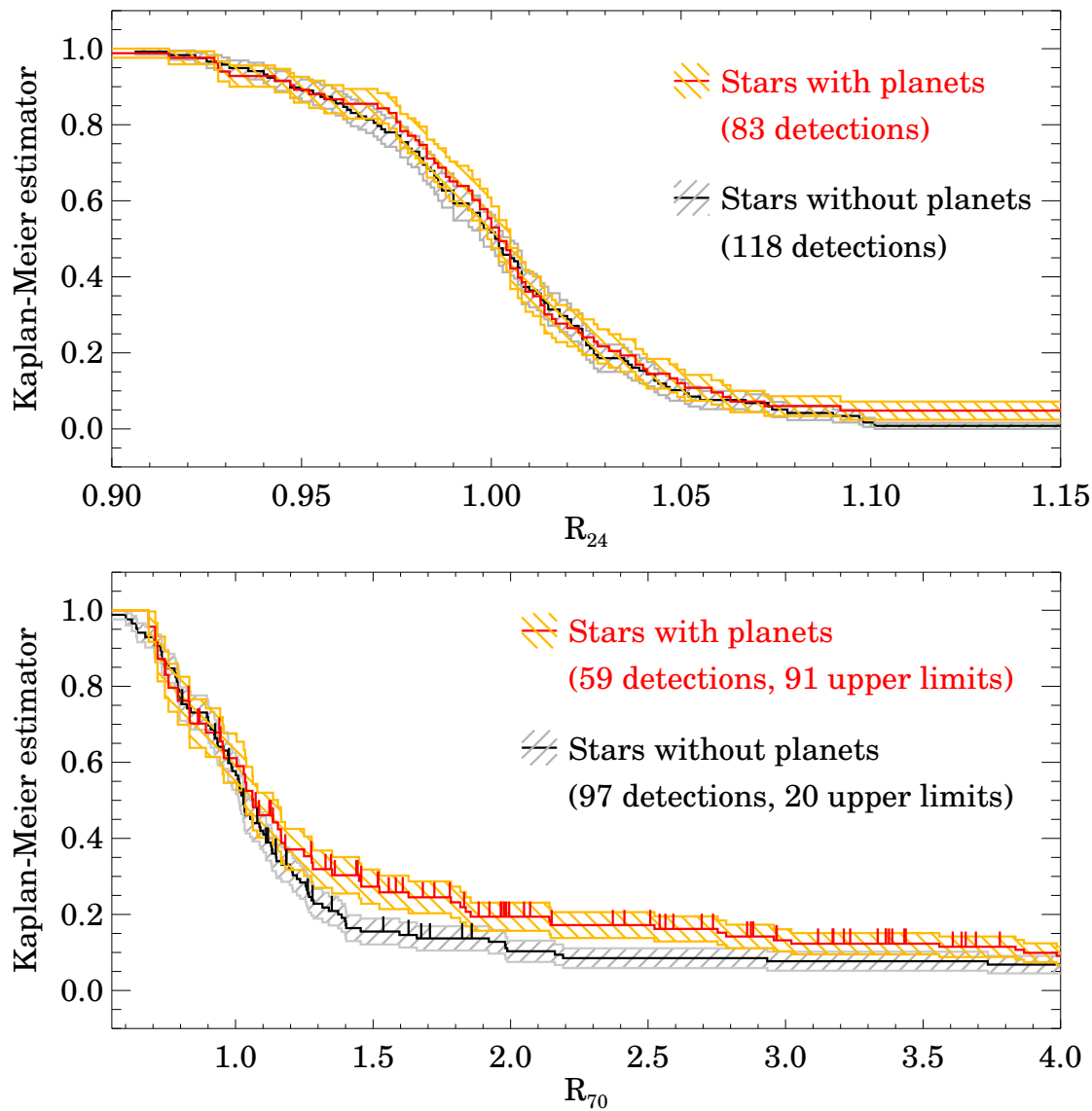}
\end{center}
\caption{Kaplan-Meier estimators for our two samples as a function of
  observed-to-photospheric flux ratio at 24$\,\mu$m (upper panel) and
  at 70$\,\mu$m (lower panel). Upper limits are indicated by vertical
  dashes. The Kaplan-Meier estimator at a certain flux ratio gives the
  fraction of stars having a flux ratio larger than that value.
  \label{fig:km}}
\end{figure}

Figure \ref{fig:km} presents the KM estimators for our SWPs and SWOPs
samples. Calculations were done with the ASURV Rev.~1.2 package
\citep{lavalley}, which implements the methods presented in
\citet{fn}. As with the two-sample tests, the KM estimators show that
at 24$\,\mu$m there is no difference between the distribution of
excesses of SWPs and of SWOPs. At 70$\,\mu$m, however, the two KM
curves are marginally different. The largest distance between the
curves occurs at $R_{70}$=1.5. The values at this point are
0.273$\,\pm\,$0.045 and 0.155$\,\pm\,$0.034 for SWPs and SWOPs,
respectively, a 2.1$\,\sigma$ difference. Thus, SWPs tend to have
excesses slightly more often than SWOPs. Other statistical tests also
suggest that the distributions are marginally different. For example,
Gehan's generalized Wilcoxon test indicates that the probability of
the null hypothesis---that SWPs and SWOPs have the same incidence of
debris disks---is 13\%. Other tests give probabilities between 20\%
and 40\%. To increase the significance of the obtained 2.1$\,\sigma$
difference to 5$\,\sigma$ would require about five times more objects
in each sample. The mean excess at 70$\,\mu$m, as given by the area
under the KM estimator curves is 2.975$\,\pm\,$0.732 for SWPs, and
1.725$\,\pm\,$0.273 for SWOPs. Though disks around SWPs tend to be
brighter, the difference is only 1.6$\,\sigma$. The results do not
change significantly if we discard the four giant stars from the SWPs
sample.

The common way in the literature to calculate detection rates is to
divide the number of stars with excess by the number of all observed
stars (Tab.~\ref{tab:detection}). At 70$\,\mu$m this method gives a
detection rate of 15\% for both SWPs and SWOPs. However, if we were
to divide by the number of detected sources only, we would conclude
that 37\% of the SWPs have excesses, compared to 18\% of the SWOP
sample. The discrepancy of these methods is related to the presence of
upper limits, which are correctly handled in our survival analysis.

\section{Discussion}
\label{sec:discussion}

It is an open question whether the orbital parameters of the planets
correlate with the presence of debris disks \citep{chiang}. We
compared stars with and without excesses within the SWP sample. The
two-sample tests mentioned above show that the planets' semimajor axes
follow the same distribution in the two groups, while eccentricities
and masses are slightly higher for stars having both planets and
debris disks than those without debris disks, although the differences
are not statistically significant.

The mean metallicities for our stars are [Fe/H]\,=\,0.103$\,\pm\,$0.018
and $-$0.097$\,\pm\,$0.024 for SWPs and SWOPs, respectively. As
expected \citep{santos2001, vf}, two-sample tests show the metallicity
distributions to be significantly different in our samples. To check
whether there are differences in the distribution of excesses between
metal-rich and metal-poor stars, we analyzed the SWPs and SWOPs
samples separately. For each sample, we compared R$_{70}$ between the
bottom and top thirds of the metallicity distributions, using KM
estimators. We found that metallicity does not correlate with the
infrared excess in either sample, confirming results obtained for
smaller samples \citep{greaves2006, beichman2005, bryden2006}. While
metallicity is one of the strongest predictors for the presence of
giant planets \citep{vf}, it does not predict the presence of debris
disks.

\citet{moromartin} argued that the larger metallicity of SWPs implies
that more planetesimals were formed early in those systems but that
most were expelled due to the orbital evolution of the giant
planets. The planetesimal configuration is then the same that would
occur in an average metallicity SWOPs. This implies that debris disks
are more common around metal-rich stars without planets, which is in
contrast to the observations presented here.

Around a 1\,L$_{\odot}$ solar-type star, a narrow ring of dust
particles radiating as blackbodies would have the peak of its emission
(F$_{\nu}$) at 24$\,\mu$m if located at 1.7\,AU from the star. The
gravitational influence of giant planets extends at most to the 3:1
mean motion resonance \citep[e.g.,][]{moromartin2005}. For our
sample, where the average semi-major axis is 1.4\,AU, this would be
2.9\,AU. Thus, these planets might stir the planetesimals located
within a belt similar to the asteroid belt in our solar
system. However, our results show that 24$\,\mu$m excess is rare in
both SWPs and SWOPs. Moreover, not all stars exhibiting 24$\,\mu$m
excess have dust within this region: the warmest dust present in these
systems is 60\,K dust at a radius of 25\,AU for HD\,10647
\citep{liseau}, 120\,K dust at 3\,AU for HD\,22049 \citep{backman},
245\,K dust at 1\,AU for HD\,69830 \citep{lisse}, and 87\,K dust at
9.1\,AU for HD\,166 \citep{trilling2008}. This implies that warm
debris dust is intrinsically rare in both SWPs and SWOPs, thus the
physical reason of this rarity is probably unrelated to the
planets. With so few stars with 24$\,\mu$m excess, the KM estimators
cannot rule out any difference between the SWPs and the SWOPs
samples. Our observations do not preclude that there are differences
between the two samples in the distribution of asteroid belt-type
debris disks fainter than our detection limit at 24$\,\mu$m (L$_{\rm
  dust}$ / L$_{*}$ $\approx$ 10$^{-5}$, as calculated from Eqn.~11 of
\citealt{wyatt2008}).

Debris dust peaking at 70$\,\mu$m is usually too far from the planets
considered here to be affected by their presence. Dust distances for
debris disk systems with excesses at 70$\,\mu$m are mostly larger than
4-5\,AU \citep{trilling2008}. The null hypothesis is consistent with
that measurement. \citet{trilling2008} argued that stars with planets
are typically farther and in more confused regions of the sky, and
suggest that dust around them may be more common than what detections
indicate. In our sample, the median distance to the SWPs is 35 pc,
compared to the 15 pc for the SWOPs, and two-sample tests reveal these
distances to be significantly different. However, our main analysis
quantity is the distance-independent $R$, and we explicitly include
information about sky confusion noise in the form of upper limits,
which are taken into account in the survival analysis. Therefore, our
results do not support the presence of a large population of
undiscovered debris disks around SWPs.

At 70$\,\mu$m, our detection limit (Eqn.~11 of \citealt{wyatt2008}) is
L$_{\rm dust}$ / L$_{*}$ $\approx$ 10$^{-6}$. For these bright debris
disks, the null hypothesis implies that the population of planets at
distances where the 70$\,\mu$m emission comes from is the same for
SWPs and the SWOPs. The marginally higher incidence of debris disks
around SWPs suggests that in these systems an outer planet might also
exist, which stirs up the planetesimals producing a cold debris disk.

The model by \citet{wyatt2007} for A-type stars suggests that the
70$\,\mu$m emission should be larger for SWPs than for SWOPs, due to
the larger mass of the protoplanetary disk in the former. They
speculate that a similar trend holds for Sun-like stars. We found that
the brightest debris disks in our SWPs sample are 2$-$3 times brighter
than those in the SWOPs sample, but the mean brightnesses are only
slightly higher in the SWPs sample at a 1.6$\,\sigma$ level. Our
results do not contradict with the prediction of \citet{wyatt2007},
although the difference we found is less pronounced than that expected
for A-type stars. Note that disk mass is not the only variable that
controls the likelihood of planet formation: disk lifetime,
metallicity, and surface density distribution are also important.


\section{Summary and conclusions}
\label{sec:summary}

Prior to this study, only a dozen stars were known to harbour both
planets and debris disks. Our 10 new debris disks found around
planet-bearing stars at 70$\,\mu$m doubled this number. We analyze a
sample of planet-host stars at least three times larger than any
previous studies and a carefully chosen control sample of stars
without planets that is comparable in size. We have found that the
incidence of debris disks (measured at 70$\,\mu$m) is only marginally
higher among stars with planets, than among those without. This result
suggests the possibility that---if debris production is primarily due
to stirring by planets---the planet population at large radii (larger
than what current radial velocity surveys can find) is comparable in
both samples (although there are other possibilities for debris
production, such as planetesimal self-stirring, or external
perturbations). We found that the brightness of the average debris
disk is not significantly different in the two samples. We could not
discover any clear correlation between the planets' orbital parameters
or the parent stars' metallicity and the presence of debris dust.


\acknowledgments

Acknowledgments. This work is based on observations made with the {\it
  Spitzer Space Telescope}, which is operated by the Jet Propulsion
Laboratory, California Institute of Technology under a contract with
NASA. \'A.~K.~acknowledges support from the {\it Spitzer} Visiting
Graduate Student Fellowship and from the Netherlands Organization for
Scientific Research (NWO). The authors thank L. Bal\'azs for useful
discussions about statistics and the referee for his/her
comments. This work was partly supported by the Hungarian Research
Fund OTKA K62304.

{\it Facilities:} \facility{{\it Spitzer}}.


\clearpage

\end{document}